\documentclass[prc,aps,manuscript]{revtex4}
\usepackage{amsfonts}
\usepackage{amsmath}
\usepackage{graphicx}
\usepackage{dcolumn}
\usepackage{bm}

\begin{document}

\title{Nuclear descent from the fission barrier in the presence
of long--range memory effects}
\author{ S. V. \surname{Radionov} \thanks{%
Electronic address: sergey.radionov18$gmail.com}}
\affiliation{Institute for Nuclear Research, 03680 Kyiv, Ukraine} 
\date{\today}

\begin{abstract}
We have investigated the peculiarities of
nuclear descent from a parabolic fission barrier within
a generalized Langevin equation with power--law
$f(t-t')=(|t-t'|/\tau)^{-\alpha}$ memory function.
We have observed much stronger slowing down of the
nuclear descent in the presence of long--range memory effects,
caused by the power--law memory function at $0<\alpha<1$,
than in the presence of short--range memory effects,
generated by exponential $f(t-t')={\rm exp}(-|t-t'|/\tau)$
memory function. At a specific value of the exponent
$\alpha=1/2$ of the power--law memory function, it turned
out possible to find analytically the trajectory of the descent
and demonstrate that the long--range memory effects
give rise to complex time oscillations of nuclear shape,
becoming more frequent and damped with the correlation
time $\tau$. We have found fairly long ($>10^{-20}~{\rm s}$)
times of the descent of $^{\rm 236}{\rm U}$ at the values
of the correlation time $\tau \sim [10^{-24}\div 10^{-23}]~{\rm s}$.
\end{abstract}


\maketitle

\newpage

\section{Introduction.}
\label{intro}

The formalism of Langevin equation \cite{zwan61,mori65,kubo66}
is a powerful tool under transport description of different
dynamical processes in systems of many interacting particles.
In the case of nuclear many--body systems, the Langevin
approaches have been used to describe fission \cite{abe86},
fusion \cite{frob87} and deep--inelastic processes \cite{bar85}.
All these Langevin approaches are based on separation of
nuclear degrees of freedom onto a few macroscopic (collective)
$q(t)$ and a mass of microscopic (nucleonic) modes of motion
\cite{abe96,frob98}. The latter constitutes a heat bath with
temperature $T$, exerting friction
$\kappa_0\int_0^t f(t-t')[dq/dt](t')dt'$ and random $\xi(t)$
forces on collective motion, related to each
other through the fluctuation--dissipation theorem \cite{frob98}.
In general, the friction and random forces have time non--local
character, determined by time--spreaded memory function
$f(t-t')$, which represents a complex energy flow between the
collective and nucleonic degrees of freedom \cite{kid96,kora10}.
Although there is a controversial opinion on the importance of
non--Markovian (memory) effects in the nuclear large--amplitude
collective dynamics
\cite{frob98,geg08,abe93,kora01,aick04,kora06,koabra08,kora09},
we would like to stress that all these non--Markovian studies
are only using the different versions of
exponential $f(t-t')={\rm exp}(-|t-t'|/\tau)$ memory function,
where the relaxation (correlation) time $\tau$
measures the time spread of the retarded friction force.
It has been demonstrated earlier \cite{kora01} that
the non--Markovian descent from the fission barrier,
governed by the exponential memory function, may
be significantly delayed and accompanied by
characteristic oscillations of nuclear shape.  The
memory effects there show non--monotonic dependence
on the correlation time $\tau$, i. e., in two extremes
of quite small and fairly large values of $\tau$ the
nuclear collective dynamics becomes Markovian
\cite{kora01,rako15}. In fact, such memory effects
are of short--range type as far as they are only prominent
within a narrow interval of values of the correlation time $\tau$,
which are comparable to reciprocal of the characteristic
frequency of the nuclear collective motion \cite{abe93,kora01}.

In the present study we investigate nuclear fission dynamics
at the descent from the top of fission barrier to scission point
by the help of the generalized Langevin equation with a
power--law, $f(t-t')=(|t-t'|/\tau)^{-\alpha}$, memory function.
Such a memory function has been successfully applied
under the generalized Langevin description of many
dynamical systems, exhibiting anomalous diffusion
behaviour \cite{boge90,fo94,mixi06}. The anomalous character
of the diffusion there reflects in a fractional time dependence
of the mean square displacement of the system and in a
power--law decay of its velocity autocorrelation function
\cite{mekl00,za02}.
All these remarkable features of the anomalous diffusion
process are caused by the presence of long--range
memory effects in the system`s dynamics \cite{luxu98},
i. e., the memory effects, existing over a broad range
of time scales of the system`s dynamics \cite{ca99}.

The plan of the paper is as follows. In Sect.~\ref{gleom}, we set
in the generalized Langevin equation of motion for a nuclear
shape variable and give its solution in terms of the Laplace
transform of the memory function $f(t-t')$. Sect.~\ref{emk} is
devoted to discussion of the time evolution of trajectory of
nuclear descent $q(t)$ in the presence of memory effects,
caused by the exponential memory function
$f(t-t')={\rm exp}(-|t-t'|/\tau)$. In Sect.~\ref{plmk}, we give
an analytical solution of the generalized Langevin equation
of motion with the power--law memory function
$f(t-t')=(|t-t'|/\tau)^{-\alpha}$ at specific value of the
exponent $\alpha=1/2$. In Sect.~\ref{tod},
we present results of calculations of times of the
nuclear descent. Finally, summary and main conclusions
are given in Summary.

\section{Generalized Langevin equation of motion}
\label{gleom}

We start by postulating a generalized Langevin equation of
motion for a nuclear shape variable $q(t)$,
\begin{equation}
M(q)\frac{d^2q(t)}{dt^2}=-\frac{1}{2}\frac{\partial M(q)}{\partial q}
\left(\frac{dq(t)}{dt}\right)^2-\frac{\partial {\rm E_{pot}}(q)}{\partial q}-
\kappa_0\int_0^t f(t-t')\frac{dq(t')}{dt'}dt'+\xi(t),
\label{Lang}
\end{equation}
Here, $q(t)$ is measured in units of the radius $R_0$ of
equal--volume spherical nucleus,  $M(q)$ is a collective mass
parameter, ${\rm E_{pot}}(q)$ is a collective potential energy and
$\kappa_0$ is the strength of the retarded friction force related
to the random force $\xi(t)$ through the fluctuation--dissipation theorem,
\begin{equation}
\langle \xi(t)\xi(t') \rangle =T \kappa_0 f(t-t').
\label{fdt}
\end{equation}
In the latter equation, $T$ is a constant nuclear temperature and
the ensemble averaging $\langle...\rangle$ is performed over all
random realizations of the Gaussian stationary process $\xi(t)$.
The memory function $f(t-t')$ in Eqs.~(\ref{Lang}) and (\ref{fdt}) is
assumed to be a decaying function of its argument $x \equiv |t-t'|/\tau$,
\begin{equation}
f(x) \to 0,~~x \to \infty,
\label{fdecay}
\end{equation}
where a correlation time, $\tau$, measures the time spread of
the retarded friction force and defines an effective interval of
time within which values $\xi(t)$ and $\xi(t')$ of the random
force at different moments of time $t$ and $t'$ correlate with
each other.

Considering a nuclear descent from a top, $q=q_b$,
of fission barrier to some scission $q=q_{sc}$,  we
approximate the potential energy ${\rm E_{pot}}(q)$ by the
inverted parabolic dependence on $q$,
\begin{eqnarray}
{\rm E_{pot}}(q)=V_b-(M_B\omega_b^2/2)(q-q_b)^2.
\label{Epot}
\end{eqnarray}
Here, $V_b$ is the height of fission barrier and $\omega_b$
is a frequency parameter, defining the nuclear stiffness at
$q=q_b$,
\begin{equation}
\omega_b=\sqrt{\frac{1}{M_b}
\left|\frac{d^2 E_{\rm pot}(q)}{dq^2}\right|_{q=q_b}},~~~
M_b=M(q=q_b).
\label{omegab}
\end{equation}
For the mass parameter $M(q)$ it is adopted a hydrodynamical
value,
\begin{equation}
M(q)=\frac{1}{5}Am_0R_0^2\left(1+\frac{1}{2q^3}\right),
\label{M}
\end{equation}
where $A$ is the nuclear mass number, $m_0$ is nucleon mass
and $R_0=r_0 A^{1/3}$ is the radius of equal volume sphere.
In Fig.~1, we showed the potential energy ${\rm E_{pot}}(q)$
(\ref{Epot}) for the following set of parameters \cite{nisi84}:
\begin{equation}
A=236,~~V_b=8~{\rm MeV},~~q_b=1.6,~~
\hbar\omega_b=1.16~{\rm MeV}.
\label{param}
\end{equation}

\begin{figure}[t]
\centering
\includegraphics[width=8cm,height=6cm]{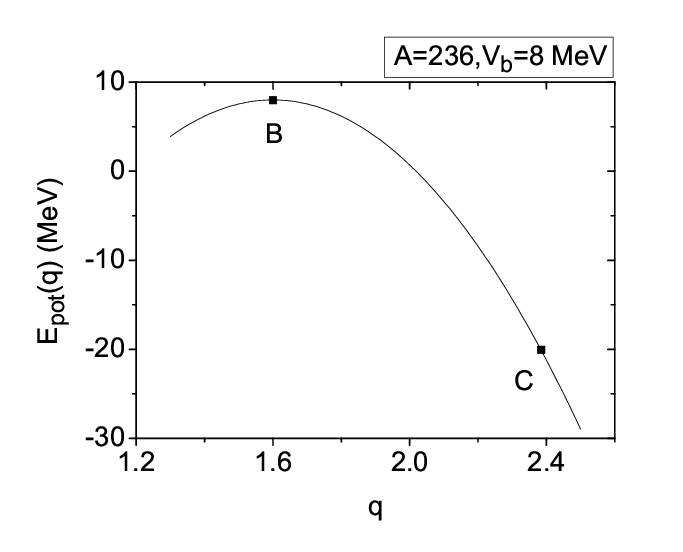}
\caption{Schematic representation of the parabolic fission
barrier (\ref{Epot}), (\ref{param}) of $^{236}{\rm U}$ with
saddle point $B$ and scission point $C$. }
\label{Fig1}
\end{figure}

\subsection{General solution of the non--Markovian dynamics}

To get analytical solution of the Langevin equation of motion
(\ref{Lang}), we linearize it with respect to a displacement,
$\Delta q(t) \equiv q(t)-q_b$,
\begin{equation}
\frac{d^2\Delta q(t)}{dt^2}=\omega^2_b\Delta q(t)-(\kappa_0/M_b)
\int_0^t f(t-t')\frac{d\Delta q(t')}{dt'}dt'+(1/M_b)\xi(t),
\label{Deltaqeq}
\end{equation}
with $M_b$ given by Eq.~(\ref{omegab}).
The general solution of Eq.~(\ref{Deltaqeq}), subject to the
initial conditions
\begin{equation}
\Delta q(t=0)=0,~~~[d \Delta q/dt](t=0)=v_0>0,
\label{initcond}
\end{equation}
can be written as
\begin{equation}
\Delta q(t)=B(t)v_0+(1/M_b)\int_0^t B(t-t')\xi(t')dt',
\label{gensol}
\end{equation}
where $v_0$ is the initial velocity of a nucleus, and $B(t)$
is a solution of the homogeneous equation,
\begin{eqnarray}
\frac{d^2 B(t)}{dt^2}=\omega^2_b B(t)-(\kappa_0/M_b)
\int_0^t f(t-t')\frac{dB(t')}{dt'}dt',
\nonumber\\
B(t=0)=0,~~~[dB/dt](t=0)=1.
\label{Bt}
\end{eqnarray}
The linear integro--differential equation (\ref{Bt}) can be solved
via the Laplace transformation method as
\begin{equation}
\tilde{B}(s)=\frac{1}{s^2+(\kappa_0/M_b)s\tilde{f}(s)-\omega_b^2},
\label{tildeBs}
\end{equation}
where $\tilde{B}(s)$ and $\tilde{f}(s)$ are the Laplace transforms
of the solution function $B(t)$ and the memory function $f(t)$,
respectively. If the denominator of the rational expression
(\ref{tildeBs}) has in total $N$ zeros such that
\begin{equation}
{\rm Re}(s_1) \geq {\rm Re}(s_2) \geq ... \geq {\rm Re}(s_N),
\label{order}
\end{equation}
then the first zero $s_1$ will define the long--time behaviour of the solution
function $B(t)$,
\begin{equation}
B(t) \sim {\rm e}^{s_1 t},~~~t\to \infty.
\label{longtime}
\end{equation}

\bigskip

\section{Exponential memory function}
\label{emk}

We first consider an exponential,
\begin{equation}
f(t-t')={\rm exp}\left(-\frac{|t-t'|}{\tau}\right),
\label{fexp}
\end{equation}
memory function, which recovers two Markovian limits
of the nuclear generalized Langevin dynamics (\ref{Lang}).
As it follows from Eq.~(\ref{Bt}), the characteristic time scale
of the solution function $B(t)$ variations is $1/\omega_b$ and
in the limit $\tau<<1/\omega_b$, the time integral in (\ref{Bt})
can be evaluated by parts,
\begin{equation}
-\kappa_0\int_0^t {\rm exp}\left(-\frac{|t-t'|}{\tau}\right)\frac{dB(t')}{dt'}dt'
\approx -\kappa_0\tau\frac{dB(t)}{dt},
\label{tau0}
\end{equation}
giving rise to the appearance of a Markovian friction term with
a friction coefficient, $\kappa_0\tau$, proportional to the correlation
time $\tau$. The corresponding solution function $B(t)$ is given by
\begin{equation}
B(t)=\frac{1}{s_1-s_2}
\left({\rm e}^{s_1 t}-{\rm e}^{s_2 t}\right),~~~
s_{1,2}=-\frac{\kappa_0\tau}{2M_b} \pm
\sqrt{\left(\frac{\kappa_0\tau}{2M_b}\right)^2+\omega_b^2}.
\label{Btfric}
\end{equation}
In the opposite limit of $\tau>>1/\omega_b$, the time integral in Eq.~(\ref{Bt})
gives rise to a Markovian restorative term,
\begin{equation}
-\kappa_0\int_0^t {\rm exp}\left(-\frac{|t-t'|}{\tau}\right)\frac{dB(t')}{dt'}dt'
\approx -\kappa_0 B(t),
\label{tau8}
\end{equation}
and $B(t)$ may be either an exponentially growing in time,
\begin{equation}
B(t)=\frac{1}{s_1-s_2}\left({\rm e}^{s_1 t}-{\rm e}^{s_2 t}\right),~~
s_{1,2}=\pm \sqrt{\omega_b^2-\kappa_0/M_b},~~
\frac{\kappa_0}{M_b\omega_b^2}<1,
\label{Btconserv1}
\end{equation}
or, oscillatory in time,
\begin{equation}
B(t)=\frac{1}{\Omega}{\rm sin}(\Omega  t),
~~\Omega\equiv |{\rm Im}(s_1,s_2)|=\sqrt{\kappa_0/M_b-\omega_b^2},
~~\frac{\kappa_0}{M_b\omega_b^2}\geq 1.
\label{Btconserv2}
\end{equation}
In the latter case, the nuclear system $q(t)$ (\ref{gensol}), (\ref{Bt})
remains in the close vicinity of the top $q_b$ of fission barrier
(\ref{Epot}) for infinitely long time.

In general, the time integral in Eq.~(\ref{Bt}) contains both a time--irreversible
(friction) and time--reversible (restorative) components,
\begin{equation}
-\kappa_0\int_0^t {\rm exp}\left(-\frac{|t-t'|}{\tau}\right)\frac{dB(t')}{dt'}dt'
=-\gamma(t,\tau)dB(t)/dt-\kappa(t,\tau)B(t),
\label{split}
\end{equation}
described by a friction, $\gamma(t,\tau)$, and a spring,
$\kappa(t,\tau)$, coefficients, respectively.
The corresponding solution $B(t)$ is found by
substituting the Laplace transform
$\tilde{f}(s)=1/(s-1/\tau)$ of the
memory function (\ref{fexp}) into Eq.~(\ref{tildeBs}),
\begin{equation}
\tilde{B}(s)=\frac{s-1/\tau}
{s^3+(1/\tau)s^2+(\kappa_0/M_b-\omega_b^2)s-1/\tau}=
\frac{C_1}{s-s_1}+\frac{C_2}{s-s_2}+\frac{C_3}{s-s_3},
\label{decompexp}
\end{equation}
leading us to the solution function
\begin{equation}
B(t)=C_1{\rm e}^{s_1 t}+C_2{\rm e}^{s_2 t}+C_3{\rm e}^{s_3 t}.
\label{Btexp}
\end{equation}
Here, $C_1,C_2,C_3$ are constants, defined by the initial conditions
(\ref{Bt}),
\begin{equation}
C_i=(s_i+1/\tau)\prod_{j=1(j\neq i)}^{3}\frac{1}{(s_i-s_j)},~~~i=\overline{1,3}
\label{Ciexp}
\end{equation}
and $s_1,s_2,s_3$ are three roots of the cubic secular equation:
\begin{equation}
(s/\omega_b)^3+\frac{1}{\omega_b\tau}(s/\omega_b)^2+
\left(\frac{\kappa_0}{M_b\omega_b^2}-1\right)(s/\omega_b)-
\frac{1}{\omega_b\tau}=0.
\label{seceqexp}
\end{equation}
This equation always has one real positive root, $s_1>0$,
while the other two roots $s_2$ and $s_3$ may be either both
real and negative or complex conjugated. In the latter case,
the memory effects in the dynamics (\ref{Bt}) are quite
prominent and the solution function (\ref{Btexp}) can be
rewritten as
\begin{equation}
B(t)=C_1{\rm e}^{s_1 t}+\left(C_{+}{\rm cos}(\Omega t)+
C_{-}{\rm sin}(\Omega t)\right){\rm e}^{-\Gamma t},
\label{23terms}
\end{equation}
where
\begin{equation}
\Omega=|{\rm Im}(s_2,s_3)|,~~\Gamma=|{\rm Re}(s_2,s_3)|,~~
C_{\pm}=C_2 \pm C_3.
\label{OmegaGammaexp}
\end{equation}

\section{Power--law memory function}
\label{plmk}

Now, we are going to investigate the peculiarities of the
the non--Markovian dynamics (\ref{Bt}), governed by
a power--law,
\begin{equation}
f(t-t')=(|t-t'|/\tau)^{-\alpha},
\label{fpower}
\end{equation}
memory function. In this case, the dynamics (\ref{Bt})
remains {\it essentially} non--Markovian at all positive values
of an exponent, $\alpha$, and of a parameter,
\begin{equation}
\rho_{\alpha}=
\frac{\kappa_0}{M_b\omega_b^2}(\omega_b\tau)^{\alpha}.
\label{rho}
\end{equation}
At each value of the exponent $\alpha$, $\rho_{\alpha}$ is
a dimensionless combination of the strength $\kappa_0$
and time--spread $\tau$ of the retarded friction force in
Eq.~(\ref{Bt}). In the sequel, we fix the value of the strength
$\kappa_0$, by taking it from the Fermi--liquid model
calculations of nuclear fission dynamics \cite{kora09},
$\kappa_0/(M_b\omega_b^2)=42$.

For the power--law memory function $f(t-t')=(|t-t'|/\tau)^{-\alpha}$
($\tilde{f}(s)=\tau^{\alpha}\Gamma(1-\alpha)/s^{1-\alpha}$),
the long--time rate $s_1$ (\ref{longtime}) of the solution
function $B(t)$ is determined as the largest positive root
of the secular equation
\begin{equation}
(s/\omega_b)^2+\rho_{\alpha}\Gamma(1-\alpha)
(s/\omega_b)^{\alpha}-1=0,
\label{seceqpower}
\end{equation}
where $\Gamma(x)$ is the gamma function and where
Eq.~(\ref{tildeBs}) was used. In Fig.~2, we showed by
solid lines the largest positive root $s_1/\omega_b$ of
the secular equation (\ref{seceqpower}) as a function of
the correlation time $\omega_b\tau$ and at several
values of the exponent $\alpha$ (\ref{fpower}),
$\alpha=1/4,1/2$ and $3/4$. For comparison, we also
showed by dashed line in Fig.~2  the largest positive root
$s_1/\omega_b$ of the cubic secular equation (\ref{seceqexp}),
corresponding to the exponential memory function (\ref{fexp}).

\begin{figure}[t]
\centering
\includegraphics[width=10cm,height=8cm]{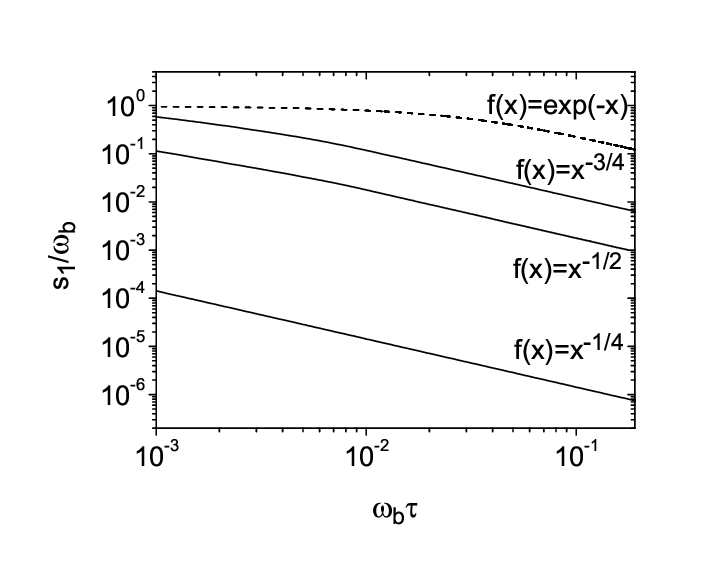}
\caption{The largest positive root $s_1/\omega_b$
of the secular equation (\ref{seceqpower}) are shown
by solid lines as a function of the correlation time
$\omega_b\tau$ and at several values of the exponent
$\alpha$ of the power--law memory function $f(x)=x^{-\alpha}$.
Dashed line is the largest positive root $s_1/\omega_b$ of
the cubic secular equation (\ref{seceqexp}), corresponding
to the exponential memory function $f(x)={\rm exp}(-x)$.}
\label{Fig2}
\end{figure}

It is seen from the Figure that the non--Markovian nuclear descent
(\ref{gensol})--(\ref{tildeBs}), measured in terms of the long--time
rate $s_1/\omega_b$ (\ref{longtime}), undergoes much slower
in the presence of the long--range $f(x)=x^{-\alpha}$ than the
short--range $f(x)={\rm exp}(-x)$ memory effects.
Such a significant (by few orders of magnitude)
slowing down of the nuclear drift is observed even at
extremely small correlation times $\omega_b\tau \sim 10^{-2}$,
at which the non--Markovian dynamics (\ref{Bt}),
subject to the exponential $f(x)={\rm exp}(-x)$ memory
function, reduces to the Markovian friction limit (\ref{tau0}).
The difference in the values of the long--time rate $s_1/\omega_b$,
calculated with the power--law  $f(x)=x^{-\alpha}$ and
exponential  $f(x)={\rm exp}(-x)$ memory functions, becomes
enormous at $\alpha \to 0$ as far as we approach
the Markovian restorative limit (\ref{tau8})
of the non--Markovian dynamics (\ref{Bt}). The stopping
of the nuclear descent here is mostly caused by the time--reversible
restorative part (\ref{split}) of the retarded friction
force in Eq.~(\ref{Bt}).

As we seen from Fig.~2, an exponentially unstable mode
of motion, ${\rm exp}^{s_1 t}$, of the solution function $B(t)$
is significantly suppressed ($s_1/\omega_b \sim [10^{-6} \div 10^{-2}]$)
and one has to define an entire time evolution of $B(t)$.
With that purpose, we restrict ourselves by considering
a particular case of $\alpha=1/2$,
for which we have found a clear analytical solution. At  $\alpha=1/2$,
the Laplace transform (\ref{tildeBs}) of $B(t)$ is given by
\begin{equation}
\tilde{B}(s)=
\frac{1}{s^2+(\kappa_0\sqrt{\pi\tau}/M_b)\sqrt{s}-\omega_b^2}
\label{tildeB1/2}
\end{equation}
and we can formally factorize the denominator of this expression
in the following way
\begin{equation}
s^2+(\kappa_0\sqrt{\pi\tau}/M_b)\sqrt{s}-\omega_b^2
=(\sqrt{s}-\mu_1)(\sqrt{s}-\mu_2)(\sqrt{s}-\mu_3)(\sqrt{s}-\mu_4).
\label{factors}
\end{equation}
The factorization (\ref{factors}) enables us to decompose the
Laplace transform (\ref{tildeB1/2}) as
\begin{equation}
\tilde{B}(s)=\frac{C_1}{\sqrt{s}-\mu_1}+\frac{C_2}{\sqrt{s}-\mu_2}
+\frac{C_3}{\sqrt{s}-\mu_3}+\frac{C_4}{\sqrt{s}-\mu_4},
\label{decomp}
\end{equation}
which in turn gives rise to the time--dependent solution function $B(t)$,
\begin{eqnarray}
B(t)=\sum_{i=1}^4 C_i\mu_i\left(1+{\rm erf}(\mu_i\sqrt{t})\right)
{\rm e}^{\mu_i^2 t},
\label{Bt1/2}
\end{eqnarray}
with ${\rm erf}(x)$ being the error function and where
\begin{equation}
C_i=\prod_{j\neq i}^{4}\frac{1}{\mu_i-\mu_j},~~~i=\overline{1,4}.
\label{Ci}
\end{equation}
In Eqs.~(\ref{Bt1/2})--(\ref{Ci}), $\mu_1,\mu_2,\mu_3,\mu_4$
are four roots of the quartic  secular equation:
\begin{equation}
(\mu/\sqrt{\omega_b})^4+\rho_{1/2}\sqrt{\pi}(\mu/\sqrt{\omega_b})-1=0,
\label{seceq1/2}
\end{equation}
where $\rho_{1/2}$ is given by Eq.~(\ref{rho}).
According to the Viet theorem, the last equation always
has two real roots (one is positive, another is negative) and
two complex conjugated roots such that
\begin{equation}
\mu_1>0,~~\mu_2<0,~~\mu_3=\mu_4^*,~~
0<|{\rm Re}(\mu_3,\mu_4)|<|{\rm Im}(\mu_3,\mu_4)|.
\label{mu}
\end{equation}
As a result of that,
\begin{equation}
C_1\mu_1\left(1+{\rm erf}(\mu_1\sqrt{t})\right){\rm e}^{\mu_1^2 t}
\to 2C_1\mu_1{\rm e}^{\mu_1^2 t},~~~\mu_1^2 t \to \infty,
\label{1term}
\end{equation}
in Eq.~(\ref{Bt1/2}) and $\mu_1^2$ may be associated with
the long--time rate $s_1$ (\ref{longtime}) of the solution function
$B(t)$. As far as $\mu_2<0$,
\begin{equation}
C_2\mu_2\left(1+{\rm erf}(\mu_2\sqrt{t})\right){\rm e}^{\mu_2^2 t}
\to \frac{C_2}{|\mu_2|\sqrt{t}},~~~\mu_2^2 t \to \infty,
\label{2term}
\end{equation}
and this term shows a power--law  decay with time. The last two terms
in the solution function $B(t)$ (\ref{Bt1/2}) can be rewritten as
\begin{equation}
\left(C_{+}(t){\rm cos}(\Omega t)
+C_{-}(t){\rm sin}(\Omega t)\right){\rm e}^{-\Gamma t},
\label{34terms}
\end{equation}
where
\begin{eqnarray}
\Omega=|{\rm Im}(\mu_3^2,\mu_4^2)|,~~~
\Gamma=|{\rm Re}(\mu_3^2,\mu_4^2)|,~~~
\nonumber\\
C_{\pm}(t)=C_3\left(1+{\rm erf}(\mu_3\sqrt{t})\right)
\pm C_4\left(1+{\rm erf}(\mu_4\sqrt{t})\right).
\label{OmegaGamma1/2}
\end{eqnarray}
As in the case (\ref{23terms}) of the exponential
$f(x)={\rm exp}(-x)$ memory function, the using of the
power--law $f(x)=x^{-1/2}$ memory function gives
also rise to the appearance of characteristic shape
oscillations of the nuclear shape $q(t)$, see
Eqs.~(\ref{gensol}), (\ref{Bt1/2}) and (\ref{34terms}).

In Figs.~3 and 4, we compared the values of the frequency
$\Omega$ and damping rate $\Gamma$ of the characteristic
shape oscillations, produced by the long--range (\ref{fpower})
and short--range (\ref{fexp}) memory effects. Solid lines in the
Figures represent the corresponding results of the calculations
of $\Omega$ and $\Gamma$ (\ref{OmegaGamma1/2}) for the
power--law $f(x)=x^{-1/2}$ memory function  and dashed lines
give the values of the frequency and damping rate
(\ref{OmegaGammaexp}) for the exponential $f(x)={\rm exp}(-x)$
memory function. Dotted line in Fig.~3 is the frequency
of the oscillations (\ref{Btconserv2}) of the trajectory of the
descent $B(t)$ in the Markovian restorative limit (\ref{tau8}).

\begin{figure}[h]
\centering
\includegraphics[width=8cm,height=6cm]{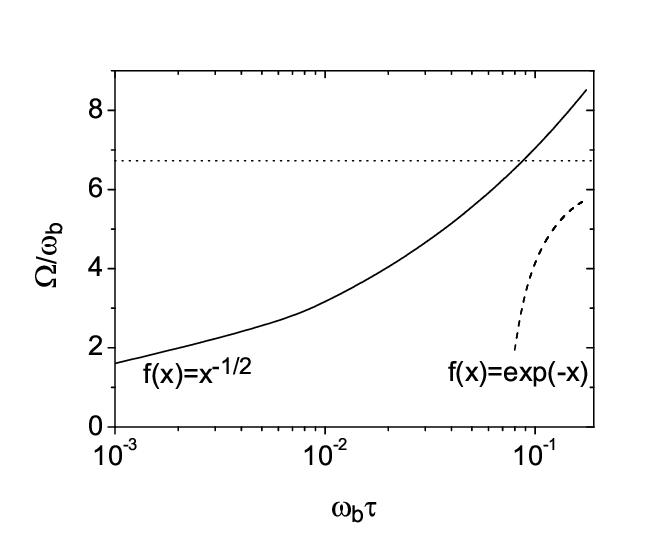}
\caption{Frequency $\Omega$ of the memory--induced
time oscillations of the trajectory of the descent $B(t)$, produced
either by the power--law $f(x)=x^{-1/2}$ (\ref{OmegaGamma1/2})
(solid line) or by the exponential $f(x)={\rm exp}(-x)$
(\ref{OmegaGammaexp}) (dashed line) memory
functions. The frequency $\Omega$ of the Markovian
restorative limit (\ref{tau8}), (\ref{Btconserv2}) is shown
by dotted line.}
\label{Fig3}
\end{figure}

\begin{figure}[h]
\centering
\includegraphics[width=8cm,height=5cm]{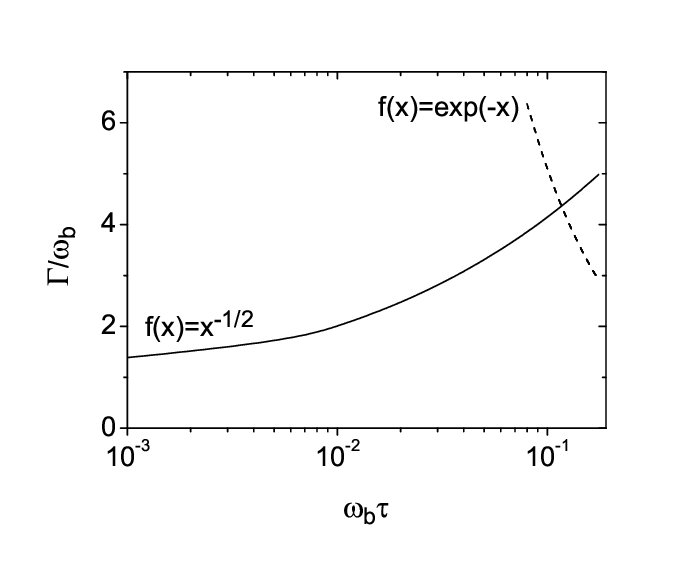}
\caption{The same as in Fig.~3 but for the damping rate
$\Gamma$ of the memory--induced time oscillations.}
\label{Fig4}
\end{figure}

From these two Figures we conclude that the typical values of the
frequency $\Omega$ and damping rate $\Gamma$ of the
memory--induced time oscillations are of comparable size.
The only thing is an opposite tendency of the damping rate
$\Gamma$ of the oscillations  with the correlation time $\tau$.
With the growth of $\tau$, the oscillations, produced by the
exponential $f(t-t')={\rm exp}(-|t-t'|/\tau)$ memory function,
become undamped, while the power--law $f(t-t')=(|t-t'|/\tau)^{-1/2}$
memory function gives rise to more damped characteristic
oscillations of the nuclear shape variable $q(t)$.

Finally, in Fig.~5 we plotted the entire solution function $B(t)$ (\ref{Bt1/2})
at two values of the correlation time $\omega_b\tau=10^{-3}$ (left panel)
and $\omega_b\tau=10^{-2}$ (right panel).

\begin{figure}[h]
\centering
\includegraphics[width=12cm,height=8cm]{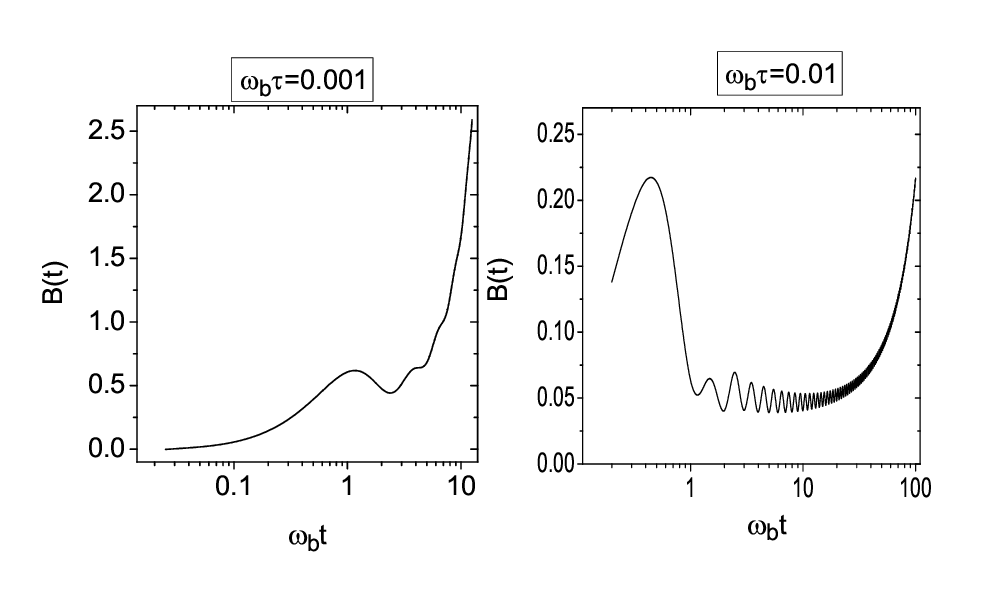}
\caption{The time evolution of the solution function $B(t)$
(\ref{Bt1/2}) at the correlation times $\omega_b\tau=10^{-3}$
and $\omega_b\tau=10^{-2}$.}
\label{Fig5}
\end{figure}

At both these values of the correlation time the nuclear descent
dynamics (\ref{Bt}), governed by the exponential ${\rm exp}(-x)$
memory function, is Markovian and the corresponding trajectory
of the descent $B(t)$ (\ref{Btfric}) is exponentially growing in time.
The trajectory of the non--Markovian descent $B(t)$ (\ref{Bt1/2}),
subject to the power--law $f(x)=x^{-1/2}$ memory function, changes
from almost exponentially growing at $\omega_b\tau=10^{-3}$
to oscillatory one at $\omega_b\tau=10^{-2}$. The former
can be associated with a regime of quite weak long--range
memory effects, while the latter situation can be associated
with sufficiently pronounced long--range
memory effects in the nuclear descent dynamics (\ref{Bt}).
The transition between the regimes occurs roughly at
$\omega_b\tau \approx 2 \times 10^{-3}$, implying
much smaller values of the correlation time $\tau$ than
in the case of using the short--range $f(x)={\rm exp}(-x)$
memory effects, which become prominent at $\omega_b\tau\sim 1$,
see Eqs.~(\ref{tau0}) and (\ref{tau8}).

\section{Time of descent}
\label{tod}

Having founded the entire solution function $B(t)$ (\ref{Bt1/2}),
we are able to estimate a duration of the nuclear descent,
${\rm t_{des}}$. We define it as a time of passing of the mean
nuclear trajectory $\langle q(t) \rangle =q_b+B(t)v0$ (\ref{gensol})
from the top $q=q_b$ (point $B$ in Fig.~1) of fission barrier to
the scission point $q=q_{sc}$ (point $C$ in Fig.~1), defined by
the following condition \cite{nisi84}:
\begin{equation}
{\rm E_{pot}}(q_b)-{\rm E_{pot}}(q_{sc})=-20~{\rm MeV},
\label{scission}
\end{equation}
see also Eq.~(\ref{Epot}). The initial velocity $v_0$ is determined
from the initial kinetic energy of the nuclear system \cite{nisi84},
\begin{equation}
\frac{1}{2}M_b v_0^2=\frac{\pi T}{4},
\label{v0}
\end{equation}
at the temperature $T=2~{\rm MeV}$.

In Fig.~6, we calculated the time ${\rm t_{des}}$ of the
non--Markovian nuclear descent (\ref{gensol}), (\ref{Bt}) in the
presence of either the power--law $f(x)=x^{-1/2}$ (solid line)
or exponential $f(x)={\rm exp}(-x)$ (dashed line) memory function.

\begin{figure}[h]
\centering
\includegraphics[width=10cm,height=8cm]{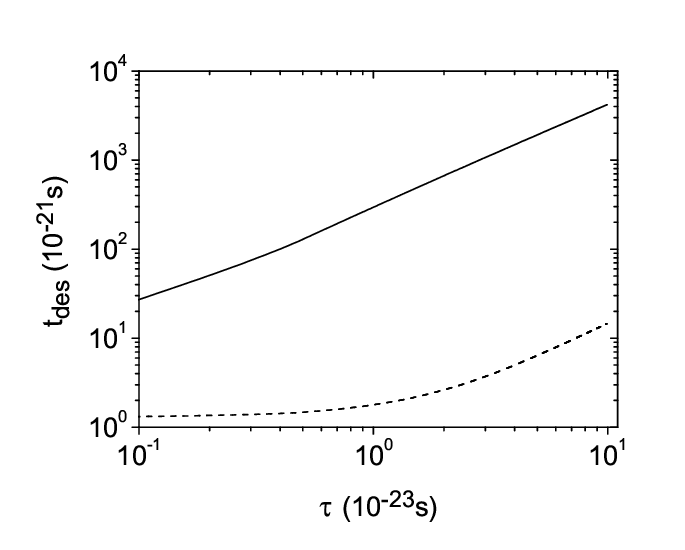}
\caption{Time of the descent ${\rm t_{des}}$
of $^{236}{\rm U}$ from the top of fission barrier (\ref{Epot})
to the scission point $q_{sc}$ (\ref{scission}). Solid and
dashed line correspond to the calculations with the
power--law $f(x)=x^{-1/2}$ and with the exponential
$f(x)={\rm exp}(-x)$ memory functions, respectively.}
\label{Fig6}
\end{figure}

It is seen from the Figure 6 that the non--Markovian dynamics
(\ref{Bt}) of the nuclear descent $q(t)$ (\ref{gensol})
takes much longer times ${\rm t_{des}}$
in the presence of the power--law $f(x)=x^{-1/2}$ than
the exponential $f(x)={\rm exp}(-x)$ memory function.
As it follows from Fig.~6, the most part of the descent time
the nuclear system $q(t)$ (\ref{gensol}) remains blocked in a close
vicinity of the top $q_b$ of fission barrier, undergoing characteristic
shape oscillations (\ref{34terms}). This quite long
oscillatory part of the descent is then followed by
a relatively short period of time during which the nuclear system $q(t)$
exponentially fast reaches the scission point $q_{sc}$ (\ref{scission}).
It is worth to point out that varying the correlation time $\tau$ from
$10^{-24}~{\rm s}$ up to $10^{-22}~{\rm s}$, one can regulate the
values of the descent time within a fairly wide interval
${\rm t_{des}} \sim [10^{-20} \div 10^{-17}]~{\rm s}]$.
Thus, at $\tau\sim 10^{-23}~{\rm s}$ the descent time
becomes comparable to typical fission time scale of
actinide nuclei $[17 \times 10^{-20}\div 40 \times 10^{-20}]~{\rm s}]$
\cite{rubc98}.

\section{Summary}

In the present study we have investigated nuclear descent
from fission barrier within the
generalized Langevin approach (\ref{Lang})--(\ref{fdt})
with the power--law memory function $f(t-t')=(|t-t'|/\tau)^{-\alpha}$.
We have linearized the Langevin equation of motion (\ref{Lang})
for a nuclear shape variable $q(t)$
in vicinity of the top of the parabolic fission barrier (\ref{Epot})
and represented its solution (\ref{gensol})--(\ref{tildeBs})
in terms of the Laplace transform $\tilde{f}(s)$ of the memory
function $f(t-t')$. In the long--time run trajectory of the descent
is an exponentially unstable, $q(t)\sim {\rm e}^{s_1 t},~~t\to\infty$,
and determined by the largest positive zero $s_1$ of the
denominator of the rational expression (\ref{tildeBs}).

We have found extremely strong suppression of the values
of $s_1$ (\ref{seceqpower}) at exponents $0<\alpha<1$
of the power--law $f(x)=x^{-\alpha}$ memory function
(solid lines in Fig.~2). On the same time, the exponential
$f(x)=exp(-x)$ memory function gives rise to much larger
(by several orders) values of $s_1$ (\ref{seceqexp})
(dashed line in Fig.~2). Thus, the explicit form of the
memory function $f(t-t')$, measuring time non--local
properties of the friction and random forces in the
generalized Langevin equation of motion
(\ref{Lang})--(\ref{fdt}), plays a crucial role
and one needs microscopic models of
$f(t-t')$

To define entire time evolution of the nuclear descent
trajectory $q(t)$, we have calculated $q(t)$ analytically
at a particular choice of the exponent $\alpha=1/2$ of
the power--law memory function (\ref{fpower}).
The obtained solution (\ref{Bt1/2}), except the exponentially
unstable $\sim {\rm e}^{s_1 t}$ (\ref{1term}), also contains
the algebraically decaying $\sim 1/\sqrt{t}$ (\ref{2term})
and the oscillating $\sim {\rm e}^{\pm i\Omega t-\Gamma t}$
(\ref{34terms}) modes of motion. The frequency $\Omega$
and damping rate $\Gamma$ (\ref{OmegaGamma1/2}) of
the oscillating modes of motion are growing
functions of the correlation time $\tau$ (Figs.~3 and 4),
implying more oscillating and damped character of the
nuclear descent (\ref{Lang}), (\ref{fpower}) with the increase
of $\tau$. By that, the nuclear system $q(t)$ remains blocked
in the close vicinity of the top of fission barrier (Fig.~5) for
sufficiently long time. We would like to comment the following
fact. On one hand, there have not been found any principal difference
in qualitative manifestation of the long--range (\ref{fpower}) and
short--range (\ref{fexp}) memory effects in the nuclear descent
dynamics (\ref{Lang}) -- both of them lead to the appearance
of characteristic oscillations of the nuclear shape, see
Eqs.~(\ref{34terms}) and (\ref{23terms}). The found difference is
rather a quantitative one -- the delay in the descent, caused by the
memory effects, takes much longer times in the case of using the
power--law memory function (\ref{fpower}). One of the reasons for
that is the presence of very slowly--decaying term $\sim 1/\sqrt{t}$
(\ref{2term}) in the solution (\ref{Bt1/2}).

We have also estimated a time of the non--Markovian descent
(\ref{Lang}) from the top $B$ of fission barrier to scission $C$
(Fig.~1). With that purpose, the time of the descent has been
calculated as a time of the first hit of the mean nuclear trajectory
$\langle q(t) \rangle$ with the scission point $q_{sc}$ (\ref{scission}).
The initial velocity of the system was taken as the mean value of
Maxwell distribution (\ref{v0}) with fixed temperature of
$T=2~{\rm MeV}$. For the nucleus $^{\rm 236}{\rm U}$,
we have found that the nuclear descent (\ref{Lang})
in the presence of the long--range memory effects
(\ref{fpower}) takes over times larger than $10^{-20}~{\rm s}$
at the range of the correlation time values
$\tau \sim [10^{-24}\div 10^{-23}]~{\rm s}$ (solid line in Fig.~6).
That implies much smaller estimation for the correlation
time $\tau$ than the one $\tau\approx 8 \times 10^{-23}~{\rm s}$,
obtained earlier in Ref.~\cite{kora09} with the exponential
memory function (\ref{fexp}) (dashed line in Fig.~6).
We also point out that at $\tau>10^{-23}~{\rm s}$ the time
of the descent becomes comparable to typical fission time
scale of actinide nuclei \cite{rubc98}.

\end{document}